\documentclass[manuscript]{aastex631}
\usepackage{amsmath}
\usepackage{bm}
\DeclareSymbolFont{lettersA}{U}{txmia}{m}{it}
\DeclareMathSymbol{\thetaup}{\mathord}{lettersA}{18}
\DeclareMathSymbol{\newtheta}{\mathord}{lettersA}{18}
\renewcommand\newtheta{\thetaup}

\DeclareSymbolFont{lettersA}{U}{txmia}{m}{it}
\DeclareMathSymbol{\deltaup}{\mathord}{lettersA}{14} 
\DeclareMathSymbol{\newdelta}{\mathord}{lettersA}{14} 
\renewcommand\newdelta{\deltaup}

\begin{document}

\title{Origin and Properties of the Near Subsonic Solar Wind Observed by Parker Solar Probe}

\author[0009-0005-3941-1514]{Wenshuai Cheng}
\affiliation{State Key Laboratory of Space Weather, National Space Science Center, Chinese Academy of \\
Sciences, Beijing, China; liuxying@swl.ac.cn}

\affiliation{University of Chinese Academy of Sciences, Beijing, China}

\author[0000-0002-3483-5909]{Ying D. Liu}
\affiliation{State Key Laboratory of Space Weather, National Space Science Center, Chinese Academy of \\
Sciences, Beijing, China; liuxying@swl.ac.cn}

\affiliation{University of Chinese Academy of Sciences, Beijing, China}

\author[0000-0002-8234-6480]{Hao Ran}
\affiliation{State Key Laboratory of Space Weather, National Space Science Center, Chinese Academy of \\
Sciences, Beijing, China; liuxying@swl.ac.cn}

\affiliation{University of Chinese Academy of Sciences, Beijing, China}

\author[0009-0004-4832-0895]{Yiming Jiao}
\affiliation{State Key Laboratory of Space Weather, National Space Science Center, Chinese Academy of \\
Sciences, Beijing, China; liuxying@swl.ac.cn}

\affiliation{University of Chinese Academy of Sciences, Beijing, China}

\author[0000-0002-7728-0085]{Michael L. Stevens}
\affiliation{Smithsonian Astrophysical Observatory, Cambridge, MA 02138, USA}

\author[0000-0002-7077-930X]{Justin C. Kasper}

\affiliation{University of Michigan, Ann Arbor, MI 48105, USA}


\begin{abstract}

We identify and examine the solar wind intervals near the sonic critical point (i.e., $M_S \sim 1$) observed by the Parker Solar Probe (PSP). The near subsonic wind intervals show similar properties: a low density, an extremely low velocity, a low proton temperature, and essentially no magnetic field deflections compared with the surrounding solar wind. The extremely low velocity is the primary contributor to the near crossing of the sonic critical point rather than the sound speed, which is roughly constant in these intervals. Source tracing with a potential field source surface (PFSS) model suggests that the near subsonic intervals all connect to the boundaries inside coronal holes. Heliospheric current sheet (HCS) and partial HCS crossings around the near subsonic intervals indicate that the near subsonic wind is a transition layer between the slow and fast wind. The above scenario is consistent with the nature of the near subsonic wind as a low Mach-number boundary layer (LMBL), which facilitates the crossing of the sonic critical point at 15-20 $R_S$. Moreover, we find a dependence of the amplitude of switchbacks on the radial sonic Mach number. Magnetic field deflections essentially disappear near the sonic critical point, which suggests that switchbacks originate from above the sonic critical point.

\end{abstract}



\section{Introduction} \label{sec:Intro}

The Parker Solar Probe (PSP) is a spacecraft that orbits closer to the Sun than any previous mission. One of its primary objectives is to investigate the mechanisms that accelerate the solar wind in the corona \citep{2016SSRv..204....7F}. When the solar wind accelerates, it undergoes two critical points, first the sonic critical point and then the Alfvén critical point \citep{1958ApJ...128..664P, 1967ApJ...148..217W}. Below these critical points lie the subsonic region and the sub-Alfvénic region, respectively. In the sub-Alfvénic region that contains the subsonic region, magnetic pressure dominates over thermal pressure (i.e., $\beta < 1$), and magnetic energy density is larger than kinetic energy density. The dominance of the magnetic field in this region makes itself a potential energy source for accelerating the plasma. \cite{2021PhRvL.127y5101K} reported PSP's first measurements of the plasma and magnetic field in the sub-Alfvénic region at a distance of about 19 solar radii from the center of the Sun. \cite{2023ApJ...944..116L} proposed a special structure originating from the boundary of a coronal hole with rapidly diverging magnetic fields, termed as a low Mach-number boundary layer (LMBL). An LMBL has an enhanced Alfvén radius, which can explain the observation of the sub-Alfvénic solar wind at a relatively large distance. A subsequent analysis has confirmed the nature of the sub-Alfvénic wind as LMBLs as an increasing number of the sub-Alfvénic intervals are sampled \citep{2024ApJ...960...42J}. 

In situ measurements of the corona near the sonic critical point are crucial for our understanding of the solar wind origin and acceleration. The location of the sonic critical point determines where the most energy deposition occurs in the solar wind and also influences whether the solar wind ultimately evolves into fast or slow wind \citep[e.g.,][]{1980JGR....85.4681L, 1982SSRv...33..161L, 2005ESASP.592..159C, 2007ApJS..171..520C}. White-light coronagraph observations and semi-empirical models accounting for the acceleration of the solar wind provide an estimate of the sonic critical radius. The location of the sonic critical point corresponding to the slow wind is at a distance of $\sim$ 5 ${R_S}$ \citep{1997ApJ...484..472S}, determined from observations of the Large Angle Spectrometric Coronagraph (LASCO) aboard the Solar and Heliospheric Observatory (SOHO) spacecraft. \cite{2003ApJ...598.1361V} found a similar location of the sonic critical point near 5 ${R_S}$ for the slow wind. The model of \cite{1995GeoRL..22.1465H} predicts a sonic critical point at 2.3 ${R_S}$ for coronal hole wind and 3.4 ${R_S}$ for denser wind structures. In situ measurements of the subsonic wind may be challenging given these values of the sonic critical radius and the final perihelion height of PSP of 9.86 ${R_S}$ from the center of the Sun \citep{2021arXiv210608450M}. However, \cite{2} reported the first measurements of the near subsonic wind at encounter 13 at a surprisingly large heliocentric distance of about 15 ${R_S}$. The near subsonic wind is characterized by an extremely low velocity and an extremely smooth radial magnetic field. They suggested that the near subsonic wind at encounter 13 is an LMBL wind by nature, which provides favorable conditions for the crossing of the sonic critical point. It is crucial to make a survey and examine whether all the near subsonic intervals possess similar characteristics and share a common origin as LMBLs, when PSP dives lower into the corona.

One of the crucial features observed by PSP is the prevalence of switchbacks in the near-Sun solar wind \citep{2019Natur.576..228K, 2019Natur.576..237B}. Switchbacks refer to magnetic field deflections with durations up to minutes, which can be larger than 90$^{\circ}$. Such radial magnetic field changes always synchronize with velocity spikes (i.e., one-sided velocity fluctuations), which suggests high Alfvénicity \citep[e.g.,][]{2014GeoRL..41..259M, 2019Natur.576..228K, 2020ApJS..246...45H, 2022ApJ...932L..13B}. The origin of switchbacks is still a debating issue. Models have been proposed to explain the generation of switchbacks, including velocity shears \citep[e.g.,][]{2006GeoRL..3314101L, 2020ApJ...902...94R,2021ApJ...909...95S}, interchange reconnection process \citep[e.g.,][]{2020ApJ...894L...4F, 2020ApJ...903....1Z, 2021A&A...650A...2D}, and expanding waves/turbulence \citep[e.g.,][]{2020ApJ...891L...2S, 2021ApJ...915...52S}. However, there is still no consensus on the origin of switchbacks. Interestingly, the amplitude of switchbacks appears to be modulated by the radial Alfvén Mach number \citep{2023ApJ...944..116L}. The amplitude of switchbacks is suppressed within the identified sub-Alfvénic intervals \citep[e.g.,][]{2021PhRvL.127y5101K, 2022ApJ...926L...1B, 2022ApJ...934L..36Z, 2024ApJ...960...42J}. \cite{2023ApJ...944..116L} found that larger deflection angles tend to occur at higher radial Alfvén Mach numbers. The relationship between the radial sonic Mach number and the amplitude of switchbacks has yet to be investigated.

In this work, we identify and examine the near subsonic intervals observed by PSP. First, we make a survey and identify all near subsonic intervals from PSP's current measurements. Second, we perform a comprehensive analysis and confirm that the near subsonic wind intervals share many of the same properties of LMBLs that are defined by \cite{2023ApJ...944..116L}. Third, we investigate the properties of the solar wind including switchbacks, as the near subsonic wind evolves into the supersonic but sub-Alfvénic wind and then to the super-Alfvénic wind. Our results suggest that switchbacks originate from above the sonic critical point, and provide additional evidence that the nature of switchbacks is outward propagating Alfvén waves. Section \ref{sec:DAM} describes the data and methodology. Details of the near subsonic intervals and their interpretations are given in Section \ref{sec:NSIAI}. We then compare the near subsonic wind with the sub-Alfvénic and super-Alfvénic wind in Section \ref{sec:Comparison}. The conclusions are summarized in Section \ref{sec:Conclusion}.

\section{Data and Methodology} \label{sec:DAM}
\subsection{PSP Data}
We use measurements provided by both the FIELDS instrument suite \citep{2016SSRv..204...49B} and the SWEAP package \citep{2016SSRv..204..131K} aboard PSP. The proton and alpha particle data are from the SWEAP ion electrostatic analyzer \citep[SPAN-I;][]{2022ApJ...938..138L}, and the magnetic field data are from the FIELDS fluxgate magnetometers. The electron density and core temperature are derived from the quasi-thermal noise (QTN) measurements \citep{2020ApJS..246...44M}. We use the electron density ${n_e}$ provided by QTN data to represent the plasma density in order to have the most accurate density determination. All the solar wind parameters from FIELDS and SWEAP are set to a cadence of 10 s. Note that the spacecraft had a high tangential velocity relative to the near subsonic wind, which enables SPAN-I to continuously measure the core of the particle velocity distribution function (VDF). Therefore, the ion data from SPAN-I are reliable in the near subsonic intervals.
\subsection{Derivation of Key Parameters}
The radial sonic Mach number is defined as $M_S = {v_R} \slash {c_S}$, where ${v_R}$ is the radial solar wind velocity and ${c_S}$ is the sound speed expressed as
\begin{equation}
    c_S = \sqrt{\frac{\gamma k_B(T_e +T_p)}{m_p}}.
    \label{cs}
\end{equation}
Here $\gamma = {5} \slash {3}$, ${k_B}$ is the Boltzmann constant, $m_p$ is the proton mass, and $T_p$ ($T_e$) is the temperature of protons (electrons), respectively. Similarly, the radial Alfvén Mach number is defined as $M_A = {v_R} \slash {v_A}$, where ${v_A}$ is the local Alfvén speed. The local Alfvén speed is written as ${v_A} = {|B|} \slash {\sqrt{\mu_0\rho}}$, where ${\mu_0}$ is the vacuum magnetic permeability, and $\rho$ is the mass density of the solar wind (consisting of the dominant component of protons). In the calculation of ${v_A}$, we use the electron number density $n_e$ from QTN to represent the plasma number density as mentioned above. In the calculation of the plasma ${\beta}$ (the ratio of thermal pressure to magnetic pressure), we take into account the contributions from electrons, protons and alphas.

To further elaborate the nature of switchbacks as outward propagating Alfvén waves, we employ the relationship between radial velocity variations and the magnetic field deflection angle derived by \cite{2023ApJ...944..116L}:
\begin{equation}
    \frac{\delta v_R}{v_A} = 1 - \cos {\theta },
    \label{one-sided}
\end{equation}
where ${\delta v_R}$ is the radial velocity variation, and $\theta$ is the magnetic field deflection angle. We follow \cite{2023ApJ...944..116L} to calculate ${\delta v_R}$ and $\theta$. A low-pass Butterworth filter with a cutoff frequency of $2 \times 10^{-4}$ Hz is used to extract ${\delta v_R}$ from the measurements. Note that $\theta$ here is the angle between the instantaneous field direction and the background field direction, which is assumed to be radial or antiradial depending on the field polarity at PSP encounter distances. A positive (negative) angle refers to a counterclockwise (clockwise) deflection in the RT plane as viewed from the north.

\subsection{Determination of Solar Source}

We use a potential field source surface (PFSS) model \citep[e.g.,][]{1969SoPh....9..131A, 2020ApJS..246...23B, 2020JOSS....5.2732S} and a ballistic propagation model to estimate the photospheric footpoints of the solar wind observed by PSP. The ballistic propagation connects the point of measurements to the source surface via an ideal Parker spiral field line, whose curvature is determined by the observed wind speed at PSP. Tracking the field lines from the source surface to the photosphere is accomplished with the PFSS model. We select the Air Force Data Assimilative Photospheric Flux Transport (ADAPT) magnetograms from the Global Oscillation Network Group (GONG) as input to the model.

We compare the model output with the Solar Dynamics Observatory \citep[SDO;][]{4} /Atmospheric Imaging Assembly \citep[AIA;][]{3} 193 $\mathring{\rm A}$ synoptic maps. For example, the darker areas in these images represent coronal holes, which can be compared with the open field regions from the PFSS model. The reconstructed coronal fields may be affected by emerging active regions \citep[e.g.,][]{2022ApJ...935...24W}. This is the case for encounters 13 and 15. Therefore, we adopt the magnetograms including active regions in order to have stable reconstructed coronal field configurations. We set appropriate source surface heights satisfying the following criteria: the height enables a reasonable alignment of the open field regions in the mapping results with observed coronal holes in AIA synoptic maps \footnote{\url{https://sun.njit.edu/coronal_holes}}, as well as a consistency in magnetic field polarities between the model results and in situ measurements. The actual heights of the source surface are given on the top of Figure \ref{fig:Trace}.

\section{Near Subsonic Intervals and Interpretations} \label{sec:NSIAI}
\subsection{Measurements at Encounters 10, 13 and 15}
\begin{figure}[ht]
    \raggedright
    \includegraphics[scale=0.78]{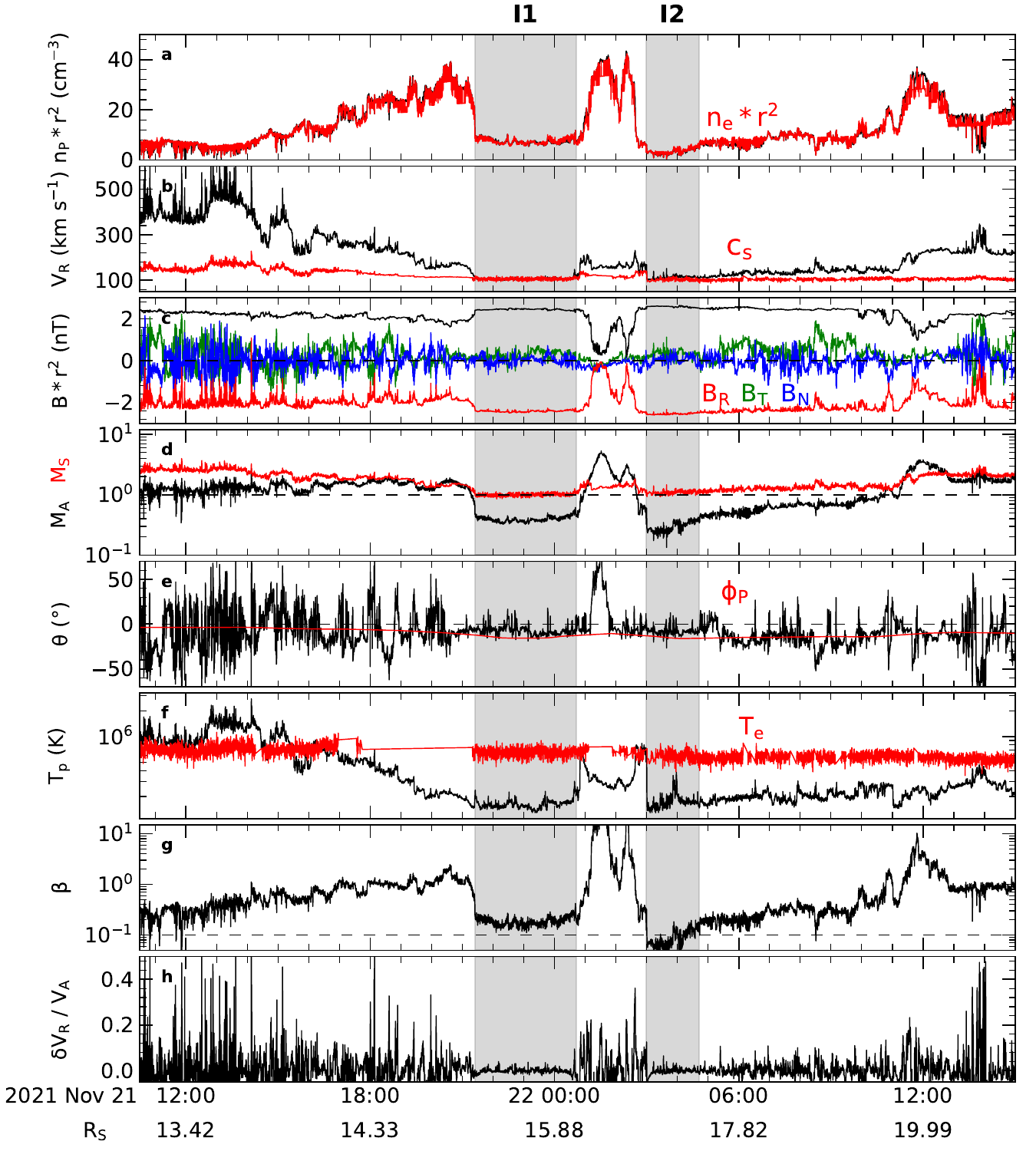}
    \caption{Overview of PSP measurements at encounter 10 from 2021 Nov 21 to 22. (a) Proton density from SPAN-I and electron density from QTN (normalized to 1 au values). (b) Proton radial velocity and the sound speed. (c) Normalized magnetic field strength and components. (d) Radial Alfvén Mach number and radial sonic Mach number. (e) Magnetic field
deflection angle and the estimated Parker spiral angle using the observed wind speed. (f) Proton temperature and electron core temperature. (g) Plasma ${\beta}$. (h) Radial velocity variation in units of local Alfvén speed. The heliocentric distance in solar radii is shown at the bottom. PSP observed two near subsonic intervals I1 and I2 at this encounter as shown by the shaded regions.}
    \label{fig:Overview_1}
\end{figure}

We examine all PSP's current measurements and only identify near subsonic intervals from measurements at encounters 10, 13 and 15 (see Table \ref{tab:table} for details about each set of measurements). Figure \ref{fig:Overview_1} shows the in situ measurements around the perihelion of encounter 10 with a transition from the fast wind to the slow wind. Inside the transition, the near subsonic intervals (I1-I2) are identified by the radial sonic Mach number $M_S \sim 1$, as shown by the shaded areas (Figure \ref{fig:Overview_1}(d)). This may represent the earliest in situ measurements of the near subsonic wind. The interruption of continuity between the two near subsonic intervals is attributed to a partial heliospheric current sheet (HCS) crossing. There is a sudden decrease in the magnetic field magnitude and an abrupt increase in the velocity, which are likely associated with reconnection exhaust in a partial HCS crossing \citep{2021ApJ...921...15C}. This scenario is consistent with the suggestion of \cite{2} that LMBLs are possibly the source of the near subsonic wind, since the location rooted in the boundary of coronal holes is indeed near the HCS. Interestingly, I1 overlaps with the sub-Alfvénic interval before the partial HCS crossing, while I2 almost begins at the same time with the other sub-Alfvénic interval after the partial HCS crossing. This suggests that the near subsonic intervals share a common source region with the sub-Alfvénic intervals as inspired by \cite{2023ApJ...944..116L}. Specifically, the near subsonic wind is also likely to be an LMBL flow. Our mapping results will further validate this speculation.

The density and magnetic field are normalized to the values at 1 AU to eliminate the effects caused by radial variations. Inside the near subsonic intervals (I1-I2), the radial component of the magnetic field is extremely smooth (Figure \ref{fig:Overview_1}(c)), and switchbacks as indicated by the spikes in $B_R$ essentially disappear. Fluctuations in $B_T$ and $B_N$ are also reduced. Compared to the surrounding supersonic but sub-Alfvénic wind and super-Alfvénic wind, I1 and I2 exhibit a smaller magnetic field deflection angle (Figure \ref{fig:Overview_1}(e)), which is essentially the background Parker spiral angle. The Parker spiral angle is estimated by the observed wind speed. The difference between the deflection angle and the Parker spiral angle can be neglected considering uncertainties from the Parker spiral geometry. The radial velocity variation also decreases correspondingly and consistently stays close to zero (Figure \ref{fig:Overview_1}(h)). Considering that the majority of the magnetic field deflection angles are below 90$^{\circ}$, we adopt the term ``ADs" as an abbreviation for Alfvénic deflections proposed by \cite{2023ApJ...944..116L} to describe the deflection properly. A reduced density (Figure \ref{fig:Overview_1}(a)) and an extremely low velocity are also obvious inside I1 and I2, which is consistent with the suggestion that PSP was likely sampling the peripheries inside coronal holes. The proton temperature is also very low corresponding to the extremely low velocity, while the electron core temperature remains roughly constant (Figure \ref{fig:Overview_1}(f)). These characteristics inside I1 and I2 cause a drop in the plasma ${\beta}$ (Figure \ref{fig:Overview_1}(g)), which suggests a lower corona for the near subsonic wind. 

Note that the decreased proton temperature in I1 and I2 tends to result in a low sound speed based on Equation (\ref{cs}) and thus inhibit the crossing of the sonic critical point (since this would increase $M_S$). However, the sound speed shows no significant changes inside and outside the near subsonic intervals. The proton temperature in I1 and I2 is up to an order of magnitude smaller than the electron core temperature, so the sound speed is primarily determined by the roughly invariant electron core temperature. Therefore, the extremely low velocity in the near subsonic intervals should be regarded as the primary contributor to the near crossing of the sonic critical point rather than the sound speed, which is nearly constant.

\begin{figure}[ht]
    \centering
    \includegraphics[scale=0.78]{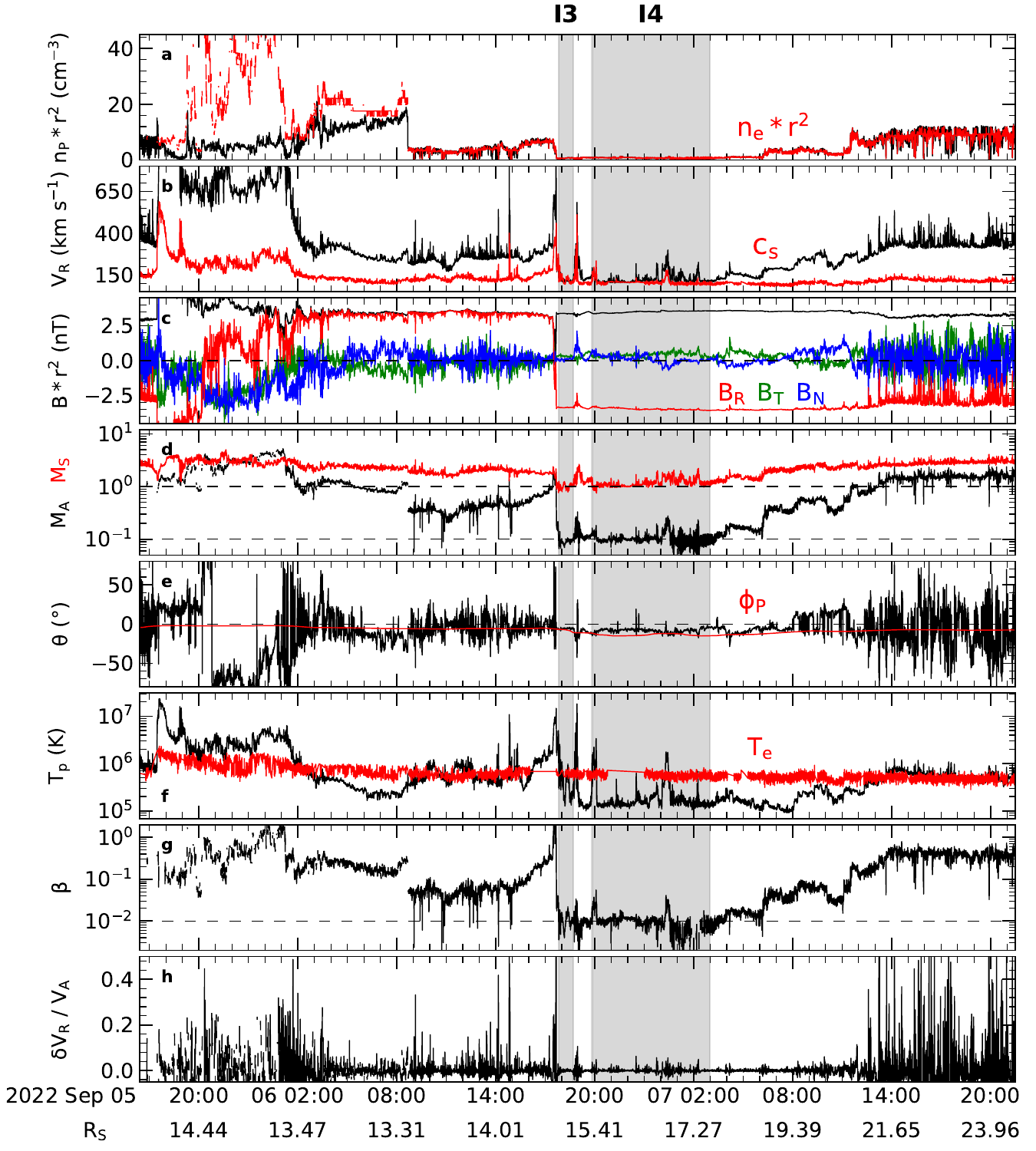}
    \caption{Overview of PSP measurements at encounter 13 from 2022 Sep 5 to 7. Similar to Figure \ref{fig:Overview_1}. Two near subsonic intervals I3 and I4 are identified. Reproduced from \cite{2}.}
    \label{fig:Overview_2}
\end{figure}

Figure \ref{fig:Overview_2} displays the in situ measurements around the perihelion of encounter 13 with a transition from a coronal mass ejection (CME) to the ambient slow wind. \cite{2} first reported the in situ measurements of the near subsonic intervals at encounter 13 with a focus on the origin and properties of the near subsonic wind. Here, we summarize the results from \cite{2} relevant to the present work. A complete HCS crossing is observed behind the CME, which occurs near 17:30 UT on September 6. The near subsonic intervals I3 and I4 are located within the sub-Alfvénic interval behind the HCS crossing, where ${M_A}$ is as low as 0.1. This observation once more supports our speculation that the near subsonic wind shares a common origin with the sub-Alfvénic wind as LMBLs. In situ observations of a large CME during encounter 13 raised some questions, such as whether the near-subsonic intervals are all induced by CME eruptions. \cite{2} suggest that the near subsonic intervals at encounter 13 are the ambient wind rather than CME leg remnants. As for the near subsonic intervals at the 10th and 15th encounters, we have checked the SOHO/LASCO CME Catalogs \footnote{\url{https://cdaw.gsfc.nasa.gov/CME_list/}} \citep{2004JGRA..109.7105Y} and found that there is no relationship between CMEs and the near subsonic intervals. Again, I3 and I4, akin to I1 and I2, display extremely smooth radial magnetic field and extremely reduced ADs (Figure \ref{fig:Overview_2}(c)). The fluctuations of $B_T$ and $B_N$ are also depressed. Again, the magnetic field deflection angle within 10$^{\circ}$ is essentially the background Parker spiral angle (Figure \ref{fig:Overview_2}(e)), and the radial velocity variation is very close to 0 (Figure \ref{fig:Overview_2}(h)). In addition, the low radial sonic Mach number $\sim 1$ in I3 and I4 is mainly caused by the extremely low velocity (Figure \ref{fig:Overview_2}(b). The overall density, velocity and proton temperature is again low in I3 and I4 in comparison with the surrounding winds, which suggests that the near subsonic wind has an origin from the peripheral region inside a coronal hole. The plasma $\beta$ is as low as 0.01 in this case, which also indicates a lower corona for the near subsonic wind.

\begin{figure}[ht]
    \centering
    \includegraphics[scale=0.78]{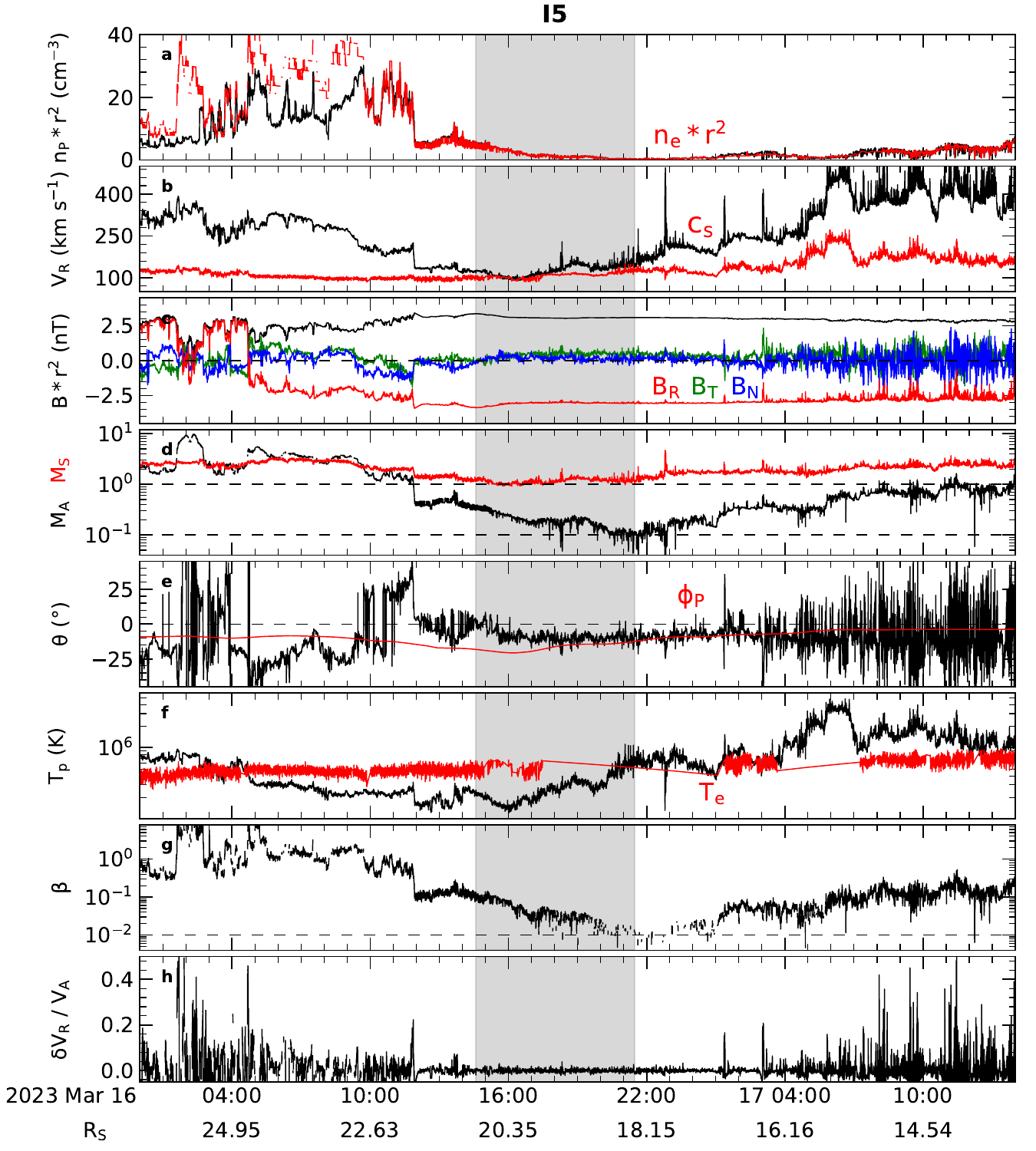}
    \caption{Overview of PSP measurements at encounter 15 from 2023 Mar 16 to 17. Similar to Figure \ref{fig:Overview_1}. One near subsonic interval I5 is identified. The identification of this interval has uncertainties because of the sparse measurements of the electron core temperature.}
    \label{fig:Overview_3}
\end{figure}

The in situ measurements around the perihelion of encounter 15 are shown in Figure \ref{fig:Overview_3} with a transition opposite to that shown in Figure \ref{fig:Overview_1}, i.e., a transition from a slow and dense wind to a fast and tenuous wind. We see a clear drop in both the radial Alfvén Mach number and the radial sonic Mach number during the transition (Figure \ref{fig:Overview_3}(d)). The slow wind before the transition is likely to originate from a streamer belt with a low speed but high plasma ${\beta}$ (see Figure \ref{fig:Overview_3}(b) and \ref{fig:Overview_3}(g)). Indeed, we observe partial HCS and HCS crossings between 02:00 UT and 05:00 UT on March 16 (Figure \ref{fig:Overview_3}(c)). Once again, the near subsonic interval I5 is within the sub-Alfvénic interval. Again, this near subsonic interval may have an origin from the borders between the slow and fast wind. It is the extremely low velocity in I5 compared with the surrounding solar wind that results in a near subsonic interval (Figure \ref{fig:Overview_3}(b)). Again, the radial component of the magnetic field is very smooth, and the fluctuations of $B_T$ and $B_N$ are reduced in I5 (Figure \ref{fig:Overview_3}(c)). The solar wind right before and behind I5 may also be the near subsonic wind given the overall low fluctuation level of the magnetic field components. The uncertainty in the extent of the near subsonic interval is caused by the sparse measurements of the electron core temperature. As in Figure \ref{fig:Overview_1} and Figure \ref{fig:Overview_2}, the near subsonic interval of I5 is also characterized by a low density, an extremely low velocity, a low proton temperature, and essentially no magnetic field deflections. Similar properties of the near subsonic intervals (I1-I5) support that these near subsonic winds share a common origin.

\subsection{Source of the Near Subsonic Intervals}
\begin{figure}[ht]
    \raggedright
    \includegraphics[scale=0.58]{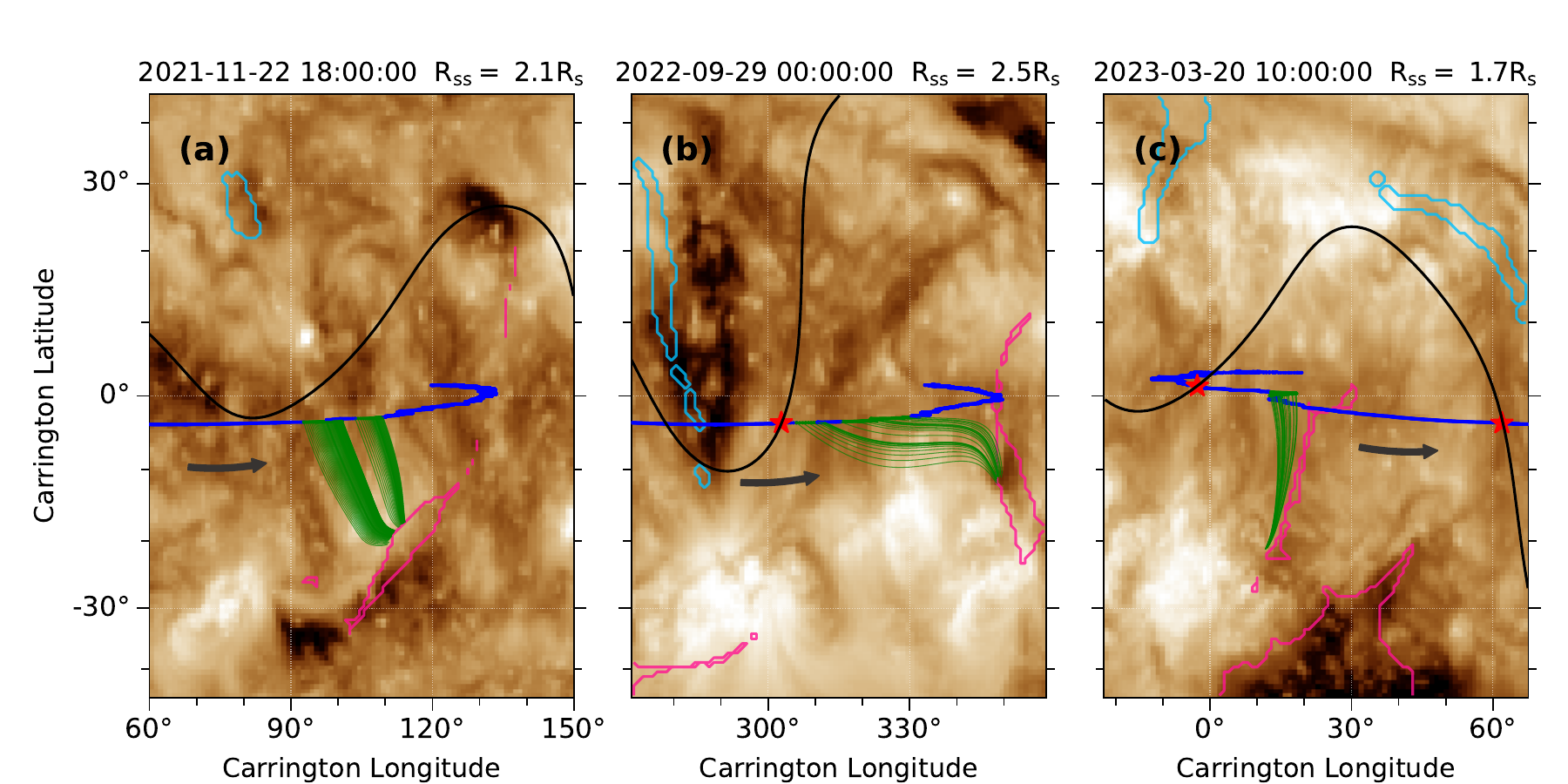}
    \caption{Magnetic mapping of the near subsonic intervals. (a) I1 and I2 at encounter 10. (b) I3 and I4 at encounter 13. (c) I5 at encounter 15. A synoptic EUV map is used as a background. The blue curve represents the trajectory of PSP projected onto the source surface, and the black arrow shows the direction of motion of the spacecraft. Pink and light blue contours indicate the open magnetic field regions in the model results with negative and positive field polarities, respectively. The black curve depicts the coronal base of the HCS. The red star indicates the location of PSP at the time of the in situ HCS crossing. The time tag of the ADAPT-GONG magnetogram and the height of the source surface used in the PFSS modeling are also shown on the top of each panel.}
    \label{fig:Trace}
\end{figure}

Figure \ref{fig:Trace} gives the magnetic mapping results for the near subsonic intervals (I1-I5). As shown in Figure \ref{fig:Trace}(a), the HCS obtained from the PFSS model is close to the PSP trajectory near a Carrington longitude of $80^{\circ}$, which is consistent with the partial HCS crossing at encounter 10. The positions of the red star in Figure \ref{fig:Trace}(b) and \ref{fig:Trace}(c) indicate in situ HCS crossings observed by PSP at encounters 13 and 15, which nearly coincide with the intersections of the HCS with the PSP trajectory. Moreover, the open magnetic field regions identified by the PFSS model are roughly consistent with the coronal holes in EUV observations at encounters 10, 13 and 15. This indicates the reliability of our mapping results. I1 and I2 from the transition between the fast and slow wind are connected to the boundaries of a low-latitude coronal hole (Figure \ref{fig:Trace}(a)). \cite{2023JGRA..12831359B} obtained similar mapping results for encounter 10. I3 and I4 at encounter 13 are connected to the boundaries of a coronal hole near the equator in Figure \ref{fig:Trace}(b). A similar mapping result is obtained by \cite{2}. As can be seen in Figure \ref{fig:Trace}(c), I5 at encounter 15 is connected to the boundary of a narrow stripe-shaded coronal hole at a low latitude once again. These are consistent with the nature of the near subsonic intervals as LMBLs \citep{2}. The origin from LMBLs explains the observed lowest velocity and reduced density inside the near subsonic intervals. It is widely accepted that there is an anticorrelation between the solar wind speed and the coronal field line expansion factor in the expansion factor model \citep[e.g.,][]{1990ApJ...355..726W}. The solar wind emanating from LMBLs along rapidly diverging open field lines would have a large expansion factor and thus a low wind speed. In addition, the LMBL wind would exhibit a low density because LMBLs root in the peripheral regions inside coronal holes. All tracing results confirm that the near subsonic winds are essentially LMBLs.

Note that these coronal holes are quite small in surface area. It is known that the smallest coronal holes are associated with unusually slow solar wind at 1 AU \citep[e.g.,][]{303}. This has been explained by stream-interaction models such as the work of \cite{2022A&A...659A.190H}. However, there is a lot of ``erosion" of high speed streams at 1 AU by stream-stream interactions. At PSP encounter distances, these interactions have not yet had time to develop, so we are getting a unique look at the pristine solar wind from these small coronal holes, and this data set provides an unprecedented way to access a flavor of the solar wind that we would never see at 1 AU.

It is likely the magnetic field configuration behind LMBLs that makes the crossing of the sonic critical point possible. In the expansion factor model, the magnetic field geometry in the low corona can alter the position of the sonic critical point by assuming that the heating rate in coronal holes depends on the local magnetic field strength \citep{2012SSRv..172..123W}. The rapidly diverging field lines inside LMBLs would result in intenser heating close to the corona base. Energy predominantly conducts downward into the transition region, which leads to a reduction in the energy available for each particle in the corona \citep{1980JGR....85.4681L, 1982SSRv...33..161L, 2012SSRv..172..123W}. Consequently, the corresponding height of the sonic critical point may increase. In light of the expansion factor model, LMBLs with overexpanded fields proposed by \cite{2023ApJ...944..116L} can be used to explain the origin of the near subsonic wind. Both the sub-Alfvénic wind and the near subsonic wind share a common source region from the boundaries inside coronal holes.

\subsection{Summary of Typical Signatures}

All the near subsonic intervals are listed in Table \ref{tab:table}. These near subsonic intervals always overlap with the sub-Alfvénic intervals as shown in Figures \ref{fig:Overview_1}-\ref{fig:Overview_3}. The average radial sonic Mach number of these intervals is 1.1, and the average radial Alfvén Mach number is 0.2. The near subsonic intervals are characterized by a low density ($\sim 2.4$ ${\rm cm^{-3}}$), an extremely low velocity ($\sim 120$ $\rm km\ s^{-1}$), a low proton temperature (${\sim 2 \times 10^5}$ K), and small deflection angles (${\sim 9^{\circ}}$). The crossing of the sonic critical point is primarily caused by the extremely low velocity in the near subsonic intervals. A deflection angle of ${\sim 9^{\circ}}$ is essentially the background Parker spiral field angle. These characteristics together with the source mapping results are indeed consistent with the nature of the near subsonic intervals as LMBLs. These near subsonic intervals are predominantly observed at around 15-20 ${R_S}$, in contrast with the values of the sonic critical radius of 2-5 ${R_S}$ in previous studies \citep[e.g.,][]{1995GeoRL..22.1465H, 1997ApJ...484..472S, 2003ApJ...598.1361V}. Therefore, we observe a particular kind of the near subsonic wind, which can be regarded as outward protrusions in the sonic critical surface just like the ``rugged" Alfvén surface suggested in previous studies \citep[e.g.,][]{2021ApJ...908L..41L, 2023ApJ...944..116L, 2023JGRA..12831359B, 2024ApJ...960...42J}. 

\section{Comparison between Near Subsonic, Sub-Alfvénic and Super-Alfvénic Winds} \label{sec:Comparison}

\begin{figure}[ht]
    \raggedright
    \includegraphics[scale=0.85]{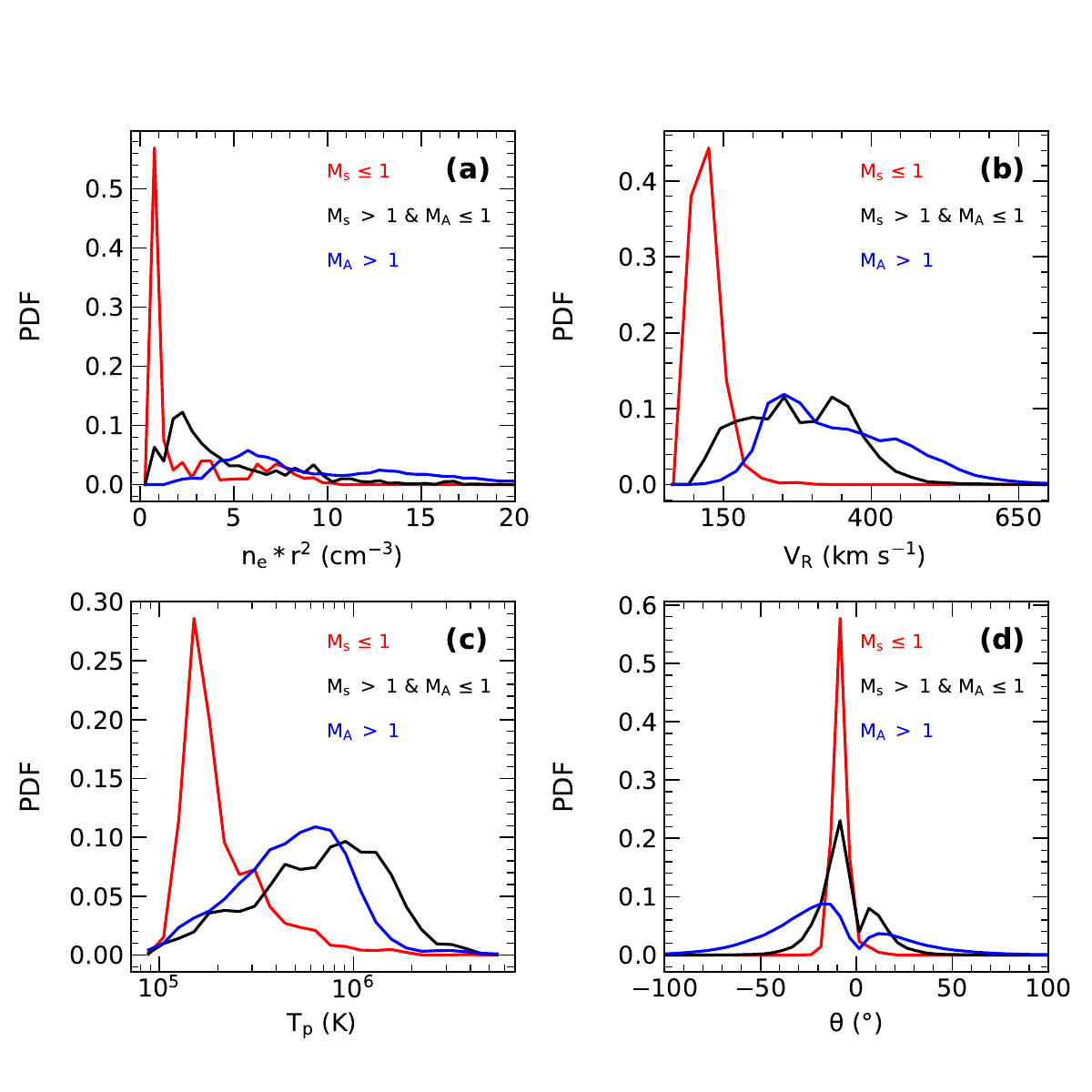}
    \caption{Probability distribution function (PDF) of normalized electron density (a), radial velocity (b), proton temperature (c), and magnetic field deflection angle (d). Here we include the near subsonic wind, supersonic but sub-Alfvénic wind and super-Alfvénic wind observed by PSP at the 8th to 15th encounters. Red is for the near subsonic wind (N = 7177), black is for the supersonic but sub-Alfvénic wind (N = 41865), and blue is for the super-Alfvénic wind (N = 662747). Here N is the number of data points used for each category.}
    \label{fig:quadruple}
\end{figure}

Figure \ref{fig:quadruple} shows the distribution of some key parameters, which indicates how the near subsonic wind differs from the supersonic but sub-Alfvénic wind and the super-Alfvénic wind. Here we use the measurements from encounters 8-15, excluding transient structures such as CMEs. The measurements cover a distance range of 13.3-59.0 $R_S$. The near subsonic wind shows characteristics that are consistent with the in situ measurements in Figures \ref{fig:Overview_1}-\ref{fig:Overview_3}: the lowest proton density, the lowest radial velocity, the lowest proton temperature, and the smallest magnetic field deflection angles compared with other types of wind. Figure \ref{fig:quadruple}(a) and Figure \ref{fig:quadruple}(b) indicate that the overall (or average) density and velocity of the super-Alfvénic wind is generally higher than that of the supersonic but sub-Alfvénic wind. This is likely because a lower density and a lower velocity are more favorable to the crossings of the Alfvén surface suggested by \cite{2021ApJ...908L..41L}, which are consistent with the typical signatures of an LMBL. The lowest velocity in the near subsonic wind may indicate an origin from the boundary of LMBLs with the most rapidly diverging open fields. As shown in Figure \ref{fig:quadruple}(c), the overall distribution of the proton temperature shifts towards lower values in the super-Alfvénic wind compared with the supersonic but sub-Alfvénic wind. This can be treated as cooling as the solar wind expands. However, the near subsonic wind has the lowest proton temperature, which is not expected. This lowest proton temperature is likely due to insufficient heating (in addition to expansion cooling) in a slow acceleration process, since the velocity is also the lowest. The asymmetry of the distribution between positive and negative magnetic field deflection angles shown in Figure \ref{fig:quadruple}(d) arises from the background Parker spiral geometry, which will introduce a clockwise (i.e., negative) deflection as viewed from the north. This asymmetry increases with the decreasing $M_A$ or $M_S$. When $M_S \leq 1$ the deflection angles are almost all negative. This indicates the increasing dominance of the magnetic field for a deeper corona. Again, the smallest deflection angles in the near subsonic wind are essentially the background Parker spiral angle. The transition from the near subsonic wind to the supersonic but sub-Alfvénic wind and then to the super-Alfvénic wind shows that the deflection angle gradually shifts towards a larger value and forms a broader distribution. This indicates a dependence of the amplitude of ADs on the radial sonic Mach number or Alfvén Mach number. 

\begin{figure}[ht]
    \centering
    \includegraphics[scale=0.88]{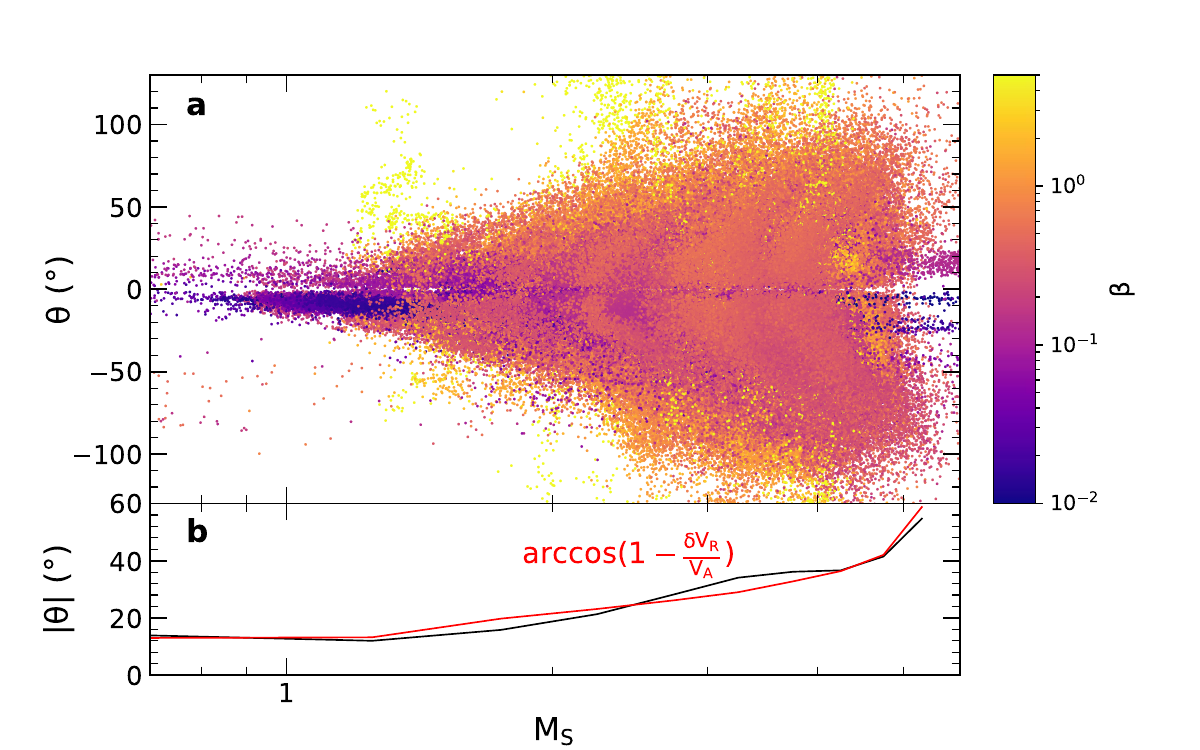}
    \caption{Magnetic field deflection angle as a function of radial sonic Mach number from PSP measurements at the 8th to 15th encounters. Colors of the dots in panel (a) represent the value of plasma $\beta$. Panel (b) shows the mean value of the absolute values of measured magnetic field deflection angles (black). The red curve is derived using Equation (\ref{one-sided}). Here we bin the data every 0.5 in $M_S$ (except a bin width of 1 for the first bin). }
    \label{fig:THETAMS}
\end{figure}

More evidence for this dependence is provided by Figure \ref{fig:THETAMS}, which shows the magnetic field deflection angle as a function of the radial sonic Mach number. The magnetic field deflection angle tends to increase with the radial sonic Mach number. An increase in the radial sonic Mach number typically corresponds to an increase in the radial Alfvén Mach number. It has been found that the deflection angle generally increases with the radial Alfvén Mach number \citep{2023ApJ...944..116L}. Hence there is a similar dependence of the amplitude of ADs on the radial sonic Mach number. A unique result is that, when the radial sonic Mach number is near 1 or below 1, the deflection angle is very small and has a flat distribution. This is essentially the background Parker spiral field angle. The plasma $\beta$ is also very small for $M_S \leq 1$ (close to 0.01), indicating a deeper corona. The remarkably low $\beta$ signifies the extreme dominance of the magnetic field in the low corona. This dominance prevents magnetic field lines from being deflected by large-scale MHD waves. Specifically, this may indicate that AD originates from above the sonic critical point. Indeed, the radial magnetic field is very smooth inside the near subsonic intervals as can be seen in Figures \ref{fig:Overview_1}-\ref{fig:Overview_3}. In addition, the fluctuations in $B_T$ and $B_N$ inside these intervals are also reduced. A similar conclusion is reached by \cite{2}. Figure \ref{fig:THETAMS} (b) shows the comparison between the absolute values of the observed deflection angles and those derived from Equation (\ref{one-sided}). They are very close to each other, which further demonstrates that ADs are essentially outward propagating Alfvén waves. Again, the average absolute value of the observed deflection angles generally increases with the radial sonic Mach number with a flat end for $M_S \sim 1$ and $M_S < 1$.

\section{Conclusions} \label{sec:Conclusion}
In this work, using PSP measurements we have identified and examined the near subsonic intervals (I1-I5) in the nascent solar wind. Key findings are revealed concerning the nature and characteristics of the near subsonic solar wind, and the comparison between the near subsonic wind, supersonic but sub-Alfvénic wind and super-Alfvénic wind. The major results are summarized as follows.
\begin{enumerate}
    \item The near subsonic intervals show similar properties: the lowest density, the lowest velocity, the lowest proton temperature, and essentially no magnetic field deflections compared with other winds. It is the lowest velocity compared with other types of wind that leads to the crossing of the sonic critical point rather than the sound speed that is roughly constant. Source mapping using the PFSS model indicates that the near subsonic intervals (I1-I5) are all connected to the boundaries inside coronal holes. In addition, HCS and partial HCS crossings around the near subsonic intervals suggest that the near subsonic wind is a transition layer between the slow and fast wind. The above scenario is consistent with the LMBL theory proposed by \cite{2023ApJ...944..116L}. 
    \item The sonic critical radius of the near subsonic intervals is 15-20 $R_S$ rather than 2-5 $R_S$ as suggested in previous studies. The near subsonic wind observed by PSP is not a typical wind. It should be regarded as an outward protrusion of the sonic critical surface. The nature of the near subsonic wind as an LMBL provides favorable conditions for the crossing of the sonic critical point. This explains that PSP observed the near subsonic wind at the distance of 15-20 $R_S$. All the results support that the near subsonic wind shares a common source region with the present sub-Alfvénic wind, and the near subsonic wind first observed by PSP is essentially an LMBL flow.
    \item We find the amplitude of ADs (our term for switchbacks to include small deflections) tends to increase with the radial sonic Mach number. ADs essentially disappear for $M_S \sim 1$ and $M_S < 1$. The radial component of the magnetic field in the near subsonic wind is very smooth, and the fluctuations in $B_T$ and $B_N$ are also reduced. The very low $\beta$ in the near subsonic wind suggests the extreme dominance of the magnetic field below the sonic critical surface, which can prevent the deflection of the magnetic field lines by large-scale MHD waves. This indicates that ADs originate from above the sonic critical point as first pointed out by \cite{2}. Our results also suggest that ADs are essentially outward propagating Alfvén waves. Therefore, Alfvén waves seem to be extremely reduced below the sonic critical point.
\end{enumerate}


The research was supported by NSFC under grant 42274201, by the Strategic Priority Research Program of the Chinese Academy of Sciences, (grant No.XDB 0560000), by the National Key R\&D Program of China (No.2021YFA0718600), and by the Specialized Research Fund for State Key Laboratories of China. We acknowledge the NASA Parker Solar Probe mission and the SWEAP and FIELDS teams for the use of data. The PFSS extrapolation is performed using the $pfsspy$ Python package \citep{2020JOSS....5.2732S}. The data used for PFSS modeling are courtesy of GONG and SDO/AIA. 

%





\bibliography{sample631}{}
\bibliographystyle{aasjournal}



\newpage

\newpage

\newpage

\newpage
\movetabledown=2in

\begin{rotatetable}

\begin{deluxetable*}{ccccccccccccc}

\tablecaption{Near Subsonic Intervals from PSP Measurements at Encounters 10, 13, and 15.}
\label{tab:table}
\tablehead{
\colhead{No.} & \colhead{Enc.} & \colhead{Start} & \colhead{duration} & \colhead{r} & \colhead{$\rm M_A$} & \colhead{$\rm M_S$} & \colhead{$\rm v_R$} & \colhead{$\rm T_p$} & \colhead{$\rm n_e \cdotp  r^{2}$}  & \colhead{$\lvert \newtheta \rvert$} 
\\
\colhead{} & \colhead{} & \colhead{(UT)} & \colhead{(hr)} &\colhead{($\rm R_S$)} & \colhead{} & \colhead{} & \colhead{($\rm km\ s^{-1}$)} & \colhead{($\rm {10^5}K$)} & \colhead{($\rm cm ^{-3}$)}  & \colhead{($\rm ^{\circ}$)} 
\\
\colhead{(1)} & \colhead{(2)} & \colhead{(3)} & \colhead{(4)} &\colhead{(5)} & \colhead{(6)} & \colhead{(7)} & \colhead{(8)} & \colhead{(9)} & \colhead{(10)} & \colhead{(11)}}

\startdata
1 & 10 & 2021-11-21 21:24:35 &  3.3 & 15.6 ${\pm}$ 0.3 & 0.39 ${\pm}$ 0.04 & 1.00 ${\pm}$ 0.05 & 106.2 ${\pm}$ 4.9 & 1.6 ${\pm}$ 0.2 & 7.5 ${\pm}$ 1.0 & 8.5 ${\pm}$ 3.7 \\
2 & 10 & 2021-11-22 02:58:55 & 1.7 & 17.1 ${\pm}$ 0.2 & 0.32 ${\pm}$ 0.05 & 1.09 ${\pm}$ 0.07 & 112.5 ${\pm}$ 8.1 & 1.8  ${\pm}$ 0.6  & 4.1 ${\pm}$ 1.0 & 8.7 ${\pm}$ 3.2 \\
3 & 13 & 2022-09-06 17:49:45 & 0.9 & 14.9 ${\pm}$ 0.1 & 0.10 ${\pm}$ 0.01 & 1.01 ${\pm}$ 0.11 & 118.2 ${\pm}$ 9.8 & 4.3  ${\pm}$ 3.0  & 0.70 ${\pm}$ 0.05 & 5.8 ${\pm}$ 0.8 \\
4 & 13 & 2022-09-06 19:50:15 & 7.1 & 16.4 ${\pm}$ 0.6 & 0.12 ${\pm}$ 0.03 & 1.20 ${\pm}$ 0.22 & 127.6 ${\pm}$ 30.5 & 2.4 ${\pm}$ 3.1 & 0.85 ${\pm}$ 0.10 & 8.4 ${\pm}$ 2.5 \\
5 & 15 & 2023-03-16 14:34:25 & 6.7 & 19.6 ${\pm}$ 0.7 & 0.19 ${\pm}$ 0.05 & 1.17 ${\pm}$ 0.13 & 125.6 ${\pm}$ 19.6 & 2.8 ${\pm}$ 1.2 & 1.40 ${\pm}$ 0.90 & 9.0 ${\pm}$ 3.5 \\
\hline
\multicolumn{2}{c}{Average} & {} & {} & 17.4 ${\pm}$ 1.8 & 0.20 ${\pm}$ 0.11 & 1.14 ${\pm}$ 0.17 & 121.6 ${\pm}$ 23.4 & 2.4 ${\pm}$ 2.2 & 2.42 ${\pm}$ 2.53 & 8.5 ${\pm}$ 3.2
\enddata
\tablecomments{Columns (1-4) stand for the interval number, encounter number, start time of the interval, and duration of the interval. Columns (5-11) refer to the mean value and standard deviation of PSP heliocentric distance, radial Alfvén Mach number, radial sonic Mach number, radial velocity, proton temperature, normalized electron density, and the absolute value of magnetic field deflection angle. Each of the near subsonic intervals is an LMBL by nature (see text).}
\end{deluxetable*}
\end{rotatetable}
\newpage

\newpage

\end{document}